\documentclass{optica-article}

\journal{opticajournal}

\setprjcopyright

\articletype{Research Article}

\usepackage{lineno}

\begin{document}

\title{Foundry manufacturing of tight-confinement, dispersion-engineered, ultralow-loss silicon nitride photonic integrated circuit}

\author{Zhichao Ye,\authormark{1} Haiyan Jia,\authormark{1} Zhangjun Huang,\authormark{1} Chen Shen,\authormark{2,3} Jinbao Long,\authormark{2,3} Baoqi Shi,\authormark{2,4} Yi-Han Luo,\authormark{2,3} Lan Gao,\authormark{2,3}\\ Wei Sun,\authormark{2} Hairun Guo,\authormark{5} Jijun He,\authormark{6} and Junqiu Liu\authormark{2,7,*}}

\address{
\authormark{1}Qaleido Photonics, Hangzhou 310000, China\\
\authormark{2}International Quantum Academy, Shenzhen 518048, China\\
\authormark{3}Shenzhen Institute for Quantum Science and Engineering, Southern University of Science and Technology, Shenzhen 518055, China\\
\authormark{4}Department of Optics and Optical Engineering, University of Science and Technology of China, Hefei 230026, China\\
\authormark{5}Key Laboratory of Specialty Fiber Optics and Optical Access Networks, Shanghai University, Shanghai 200444, China\\
\authormark{6}Key Laboratory of Radar Imaging and Microwave Photonics, Ministry of Education, Nanjing University of Aeronautics and Astronautics, Nanjing 210016, China\\
\authormark{7}Hefei National Laboratory, University of Science and Technology of China, Hefei 230088, China
}

\email{\authormark{*}liujq@iqasz.cn} 

\begin{abstract}
The foundry development of integrated photonics has revolutionized today's optical interconnect and datacenters. 
Over the last decade, we have witnessed the rising of silicon nitride (Si$_3$N$_4$) integrated photonics, which is currently transferring from laboratory research to foundry manufacturing. 
The development and transition are triggered by the ultimate need of low optical loss offered by Si$_3$N$_4$, which is beyond the reach of silicon and III-V semiconductors.  
Combined with modest Kerr nonlinearity, tight optical confinement and dispersion engineering, Si$_3$N$_4$ has today become the leading platform for linear and Kerr nonlinear photonics, and has enabled chip-scale lasers featuring ultralow noise on par with table-top fiber lasers. 
However, so far all the reported fabrication processes of tight-confinement, dispersion-engineered Si$_3$N$_4$ photonic integrated circuit (PIC) with optical loss down to few dB/m have only been developed on 4-inch or smaller wafers.  
Yet, to transfer these processes to established CMOS foundries that typically operate 6-inch or even larger wafers, challenges remain. 
In this work, we demonstrate the first foundry-standard fabrication process of Si$_3$N$_4$ PIC with only 2.6 dB/m loss, thickness above 800 nm, and near 100\% fabrication yield on 6-inch wafers.  
Such thick and ultralow-loss Si$_3$N$_4$ PIC enables low-threshold generation of soliton frequency combs. 
Merging with advanced heterogeneous integration,  active ultralow-loss Si$_3$N$_4$ integrated photonics could pave an avenue to addressing future demands in our increasingly information-driven society.
\end{abstract}

\section{Introduction}
Integrated photonics \cite{Thomson:16} enables the synthesis, processing and detection of optical signals using photonic integrated circuit (PIC).
The successful translation from laboratory research to foundry development over the past decades has established integrated photonics as a standard technology \cite{Rickman:14} deployed in high-data-rate telecommunication and datacenters \cite{Agrell:16}. 
Foundry-level manufacturing of photonic chips allows fast prototyping or mass production with high yield, high throughput, low cost, and guaranteed performance. 

Silicon (Si) and indium phosphide (InP) are two mainstream platforms of integrated photonics~\cite{Margalit:21}.  
The development of heterogeneous integration \cite{Roelkens:10, Komljenovic:16} has married these two platforms and created the first electrically pumped InP/Si lasers \cite{Fang:06, Liang:10} that have today been used for optical interconnect. 
Despite, Si and InP still have many limitations, particularly the high linear and nonlinear losses (e.g. two-photon absorption) that compromise their performance. 

To address this challenge, silicon nitride (Si$_3$N$_4$) emerges as a leading platform for low-loss integrated photonics \cite{Moss:13, Gaeta:19, Xiang:22}. 
The 5 eV bandgap of Si$_3$N$_4$ makes it transparent from ultraviolet to mid-infrared, and immune to two-photon absorption in the telecommunication band.  
Meanwhile, Si$_3$N$_4$ has a dominant Kerr nonlinearity but negligible Raman \cite{Porcel:17} and Brillouin nonlinearities \cite{Gyger:20}. 
In addition,with advanced CMOS fabrication techniques, linear optical loss down to 1 dB/m or even lower has been only achieved in Si$_3$N$_4$ \cite{Liu:21, Ye:21a, Ji:21, Bauters:11, Spencer:14, Puckett:21} among all integrated platforms. 
All these advantages have triggered the rapid development of Si$_3$N$_4$ Kerr nonlinear photonics \cite{Moss:13, Gaeta:19}, and have enabled key advances such as optical frequency comb generation \cite{Brasch:15, Xue:15, Joshi:16},  supercontinuum generation \cite{Halir:12, Zhao:15, Guo:18}, and quantum light sources \cite{Kues:19, Vaidya:20, Lu:19a}.  
In addition to ultralow loss, tight optical confinement with SiO$_2$ cladding is simultaneously required. 
Since the refractive index of Si$_3$N$_4$ ($n\approx1.99$) is modestly higher than that of SiO$_2$ ($n\approx1.45$), Si$_3$N$_4$ waveguides require sufficient thickness to achieve tight optical confinement for small mode volume and for bending radii down to 20 $\mu$m. 
In addition, while Si$_3$N$_4$ material has intrinsic normal group velocity dispersion (GVD) at telecommunication bands, Si$_3$N$_4$ waveguides with thickness above 600 nm can obtain net anomalous GVD that is required for phase matching in Kerr parametric processes \cite{Kippenberg:04, Okawachi:14}. 

While CMOS foundries have already developed standard Si$_3$N$_4$ processes \cite{Munoz:19} to fabricate PIC with typical thickness of 300 nm and loss on the order of 10 dB/m, there has not been a process to simultaneously achieve ultralow loss (e.g.  below 3 dB/m) and large thickness (above 600 nm) without crack formation.
So far, thin Si$_3$N$_4$ PIC with width above 5 $\mu$m and thickness below 100 nm \cite{Bauters:11, Spencer:14, Puckett:21} can achieve optical loss below 0.1 dB/m. 
While this process has recently become an 8-inch foundry process \cite{Jin:21}, the thin Si$_3$N$_4$ exhibits weak optical confinement and thus suffers from exaggerated bending loss with a small footprint. 
The optical mode also exhibits strong normal GVD due to the waveguide geometry. 
While coupled waveguide structures can be used to alter local GVD \cite{Kim:17, Yuan:23}, they cannot offer anomalous GVD over a wide spectral range. 
In parallel, thick Si$_3$N$_4$ PIC of ultralow loss and broadband anomalous GVD has been realized via the subtractive process \cite{Xuan:16, Ji:17, Ye:19b} and the photonic Damascene process \cite{Pfeiffer:18b, Liu:21}.
However, all these reported processes have been developed only in laboratories and have issues with transferring to foundry manufacturing.  
While high-yield, wafer-scale fabrication of Si$_3$N$_4$ PIC has achieved 1 dB/m loss and anomalous GVD in Ref. \cite{Liu:21}, the photonic Damascene process has intrinsic limitations. 
For example, the use of chemical-mechanical planarization (CMP) to remove excess Si$_3$N$_4$ can cause serious dishing effect in large-area structures (see Appendix C), thus cannot be used to fabricate elements such as arrayed waveguide gratings (AWG) and multimode interferometers (MMI).  
Meanwhile, the aspect-ratio-dependent etch effect prevents to form narrow but deep channels \cite{Liu:18}. 
In comparison, with the subtractive process \cite{Xuan:16, Ji:17, Ye:19b}, thick Si$_3$N$_4$ PIC of ultralow loss and anomalous GVD has been only achieved with electron-beam lithography (EBL) on 3- or 4-inch wafers, which are not foundry-standard. 
Meanwhile, an 8-inch foundry process has been developed in Ref. \cite{ElDirani:19}, but this process has not yet achieved optical loss below 3 dB/m. 

In this work, we overcome the above challenges and demonstrate a foundry-standard fabrication process of tight-confinement, dispersion-engineered, ultralow-loss Si$_3$N$_4$ PIC. 
The process is based on 6-inch wafers and combines deep-ultraviolet (DUV) stepper lithography \cite{Liu:21} and state-of-the-art subtractive process \cite{Ye:19b}, i.e. a DUV subtractive process. 
We have achieved a linear optical loss of 2.6 dB/m in 810-nm-thick Si$_3$N$_4$ PIC. 
Finally, we generate single soliton microcombs of 100.17 GHz and 19.975 GHz mode spacings using these devices. 

\section{Fabrication}
\begin{figure*}[t!]
\centering
\includegraphics[width=13cm]{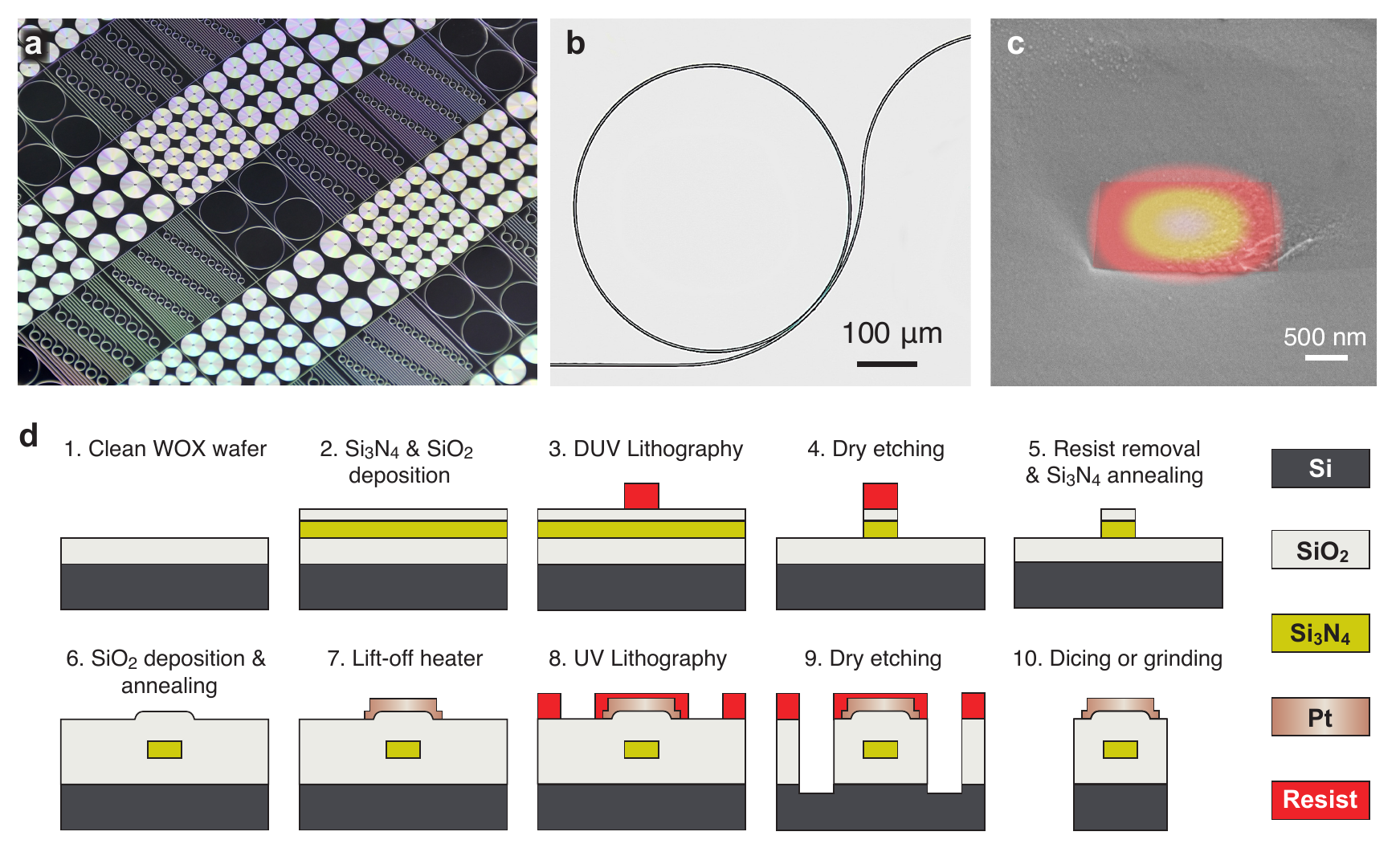}
\caption{
\textbf{Process flow and sample images of the 6-inch Si$_3$N$_4$ foundry fabrication process}. 
\textbf{a}.  
Photograph of dozens of Si$_3$N$_4$ chips on a 6-inch wafer, which contains microresonators of different FSR and meter-long spirals.  
\textbf{b}.  
Optical micrograph showing a curved bus waveguide slowly approaching a 100-GHz-FSR microring resonator. 
\textbf{c}.  
SEM image showing the Si$_3$N$_4$ waveguide core with SiO$_2$ cladding. 
The TE$_{00}$ mode is plotted, showing tight confinement in the Si$_3$N$_4$ waveguide core.  
\textbf{d}.  
The DUV subtractive process flow.  
WOX: wet oxide (SiO$_2$).
}
\label{Fig:1}
\end{figure*}

Figure \ref{Fig:1}a presents a photograph that shows dozens of Si$_3$N$_4$ chips on a 6-inch wafer, which contains microresonators of different free spectral ranges (FSR) and meter-long spirals.  
Figure \ref{Fig:1}b presents an optical micrograph that shows a curved bus waveguide slowly approaching a 100-GHz-FSR microring resonator for light coupling. 
This coupler design can increase coupling strength and ideality, which will be described later. 
Figure \ref{Fig:1}c presents a scanning electron micrograph (SEM) that shows the actual Si$_3$N$_4$ waveguide cross-section with $85^\circ$ sidewall angle and overlaid fundamental transverse-electric (TE$_{00}$) mode. 
The optical mode is tightly confined in the Si$_3$N$_4$ waveguide core with SiO$_2$ cladding, enabling dispersion engineering and small bending radii. 

The Si$_3$N$_4$ PIC is fabricated using the DUV subtractive process. 
Figure \ref{Fig:1}d shows the subtractive process flow widely used to fabricate PIC based on essentially any material, particularly Si$_3$N$_4$ \cite{Gondarenko:09, Luke:13, Xuan:16, Ji:17, Ye:19b, ElDirani:19, Wu:20, Wu:21}. 
First, a Si$_3$N$_4$ film is deposited on a clean thermal wet SiO$_2$ substrate via low-pressure chemical vapor deposition (LPCVD). 
It is well known that LPCVD Si$_3$N$_4$ films are prone to crack due to their intrinsic tensile stress (typically 1.1 to 1.4 GPa).
The film stress can be relaxed via thermal cycling during Si$_3$N$_4$ deposition in multiple layers \cite{Gondarenko:09, ElDirani:19}, yielding zero cracks during our fabrication.  
After SiO$_2$ deposition as an etch hardmask,  DUV stepper photolithography is used to expose the waveguide pattern. 
Via dry etching, the pattern is subsequently transferred from the photoresist mask to the SiO$_2$ hardmask, and then into the Si$_3$N$_4$ layer to form waveguides.  
For superior etch quality and smooth waveguide sidewall, the etchant we use is CHF$_3$ with added O$_2$ to remove fluoride-carbon polymers as etch byproducts. 

The etched substrate is thermally annealed in nitrogen atmosphere at 1200$^\circ$C to eliminate nitrogen-hydrogen bonds in Si$_3$N$_4$, which cause absorption loss.
Top SiO$_2$ cladding is then deposited on the wafer, which also requires high-temperature annealing to remove silicon-hydrogen bonds that also cause absorption loss. 
In specific cases where deuterated plasma-enhanced chemical vapor deposition (PECVD) Si$_3$N$_4$ \cite{Chiles:18, Wu:21} and SiO$_2$ \cite{Jin:20} are used, thermal annealing may not be required as these films are intrinsically hydrogen-free. 

Platinum heaters \cite{Joshi:16, Xue:16} are deposited on the substrate via an evaporator and patterned via a lift-off process.  
Due to the thick top SiO$_2$ cladding and tight optical confinement of the Si$_3$N$_4$ core, the presence of metallic heaters does not impact the optical loss of Si$_3$N$_4$ waveguides beneath. 
Afterwards, UV photolithography and deep dry etching of SiO$_2$ and Si are used to define chip size and create smooth chip facets. 
Finally, the wafer is separated into chips using dicing or backside grinding. 

\section{Characterization}
\subsection{Microresonator quality factors and loss rates}
\begin{figure*}[t!]
\centering
\includegraphics[width=13cm]{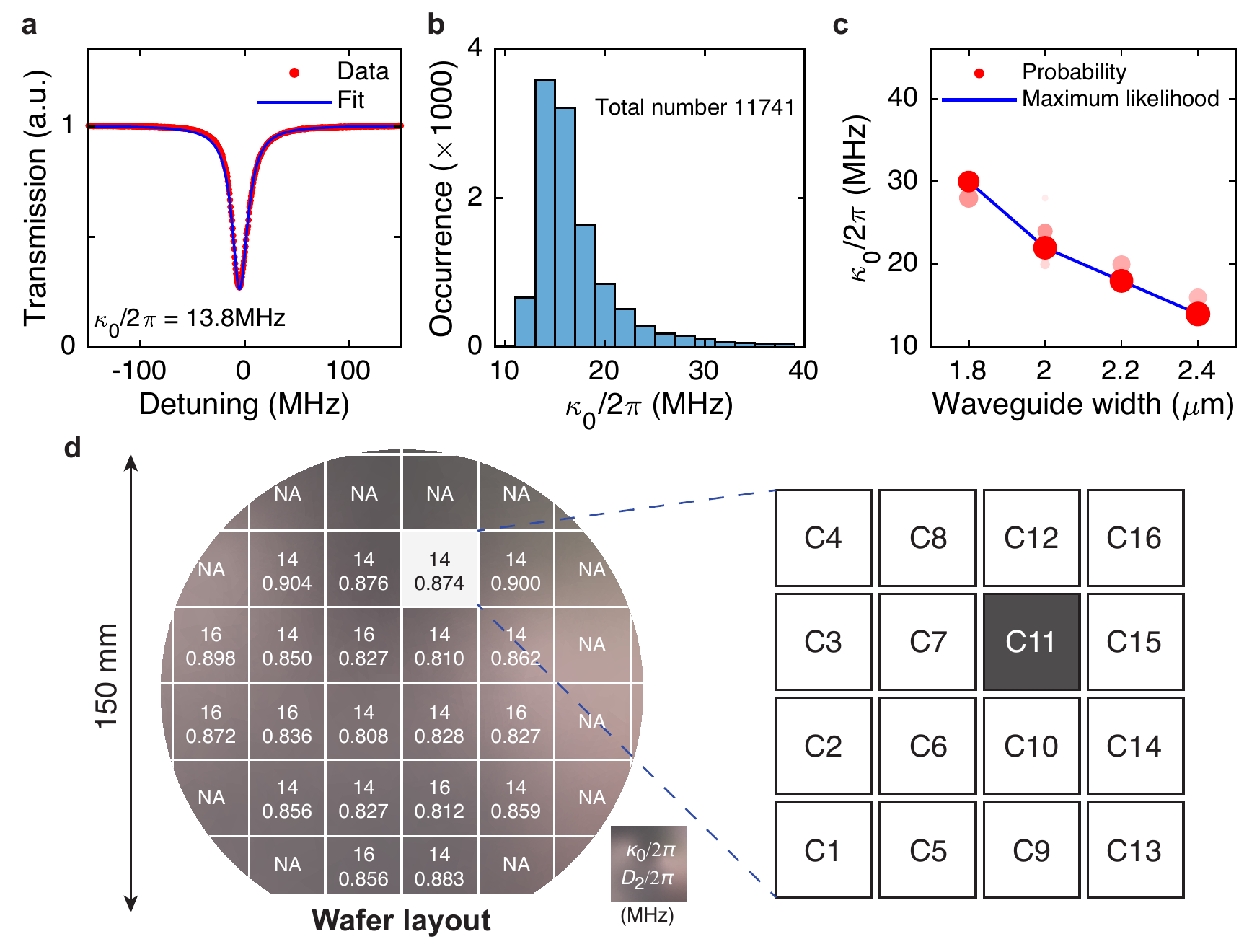}
\caption{
\textbf{Statistical loss characterization and yield analysis}. 
\textbf{a}.  
A typical TE$_{00}$ resonance profile with a Lorentzian fit, showing $\kappa_0/2\pi=13.8$ MHz and negligible mode split. 
\textbf{b}.  
Histogram of 11741 TE$_{00}$ resonances from sixty 100-GHz-FSR microresonators of 2.40 $\mu$m waveguide width, showing the most probable value of $\kappa_0/2\pi=14$ MHz and $Q_0=1.4\times10^7$.
\textbf{c}.  
Characterization of waveguide-width-dependent loss. 
Microresonators of 2.40, 2.20, 2.00 and 1.80 $\mu$m waveguide widths are characterized and compared. 
A trend of lower $\kappa_0$ with larger width is shown. 
The size and color tone of the circles indicate the probability of occurrence. 
\textbf{d}.  
Uniformity and yield analysis over the 6-inch wafer scale. 
Right: The DUV stepper reticle layout contains sixteen chips, and is uniformly exposed in discrete fields over the 6-inch wafer.
Left: The most probable values $\kappa_0/2\pi$ of the C11 chips, as well as the measured GVD parameters $D_2/2\pi$, are marked in each fields over the wafer.
NA: not applicable, due to visible defects or missing C11 chips near wafer edge.
}
\label{Fig:2}
\end{figure*}

We characterize the optical loss of Si$_3$N$_4$ PIC by measuring the resonance linewidth of Si$_3$N$_4$ microresonators.  
Light is coupled into and out of Si$_3$N$_4$ chips via lensed fibers and inverse tapers \cite{Liu:18}.  
The fiber-chip edge coupling efficiency is about 60\%. 
We use frequency-comb-assisted diode laser spectroscopy\cite{DelHaye:09} to measure resonance frequency $\omega/2\pi$ and linewidth $\kappa/2\pi$, ranging from 1480 to 1640 nm wavelength. 
The resonance's quality factor is calculated as $Q=\omega/\kappa$.
Here we study both the TE$_{00}$ and TM$_{00}$ (fundamental transverse-magnetic) modes of the 100-GHz-FSR microresonators of 810 nm thickness and 2.40 $\mu$m waveguide width. 
For each resonance fit \cite{Li:13}, the intrinsic loss $\kappa_0/2\pi$,  external coupling strength $\kappa_\text{ex}/2\pi$, and the total (loaded) linewidth $\kappa/2\pi=(\kappa_0+\kappa_\text{ex})/2\pi$, are extracted. 

Figure \ref{Fig:2}a shows a typical TE$_{00}$ resonance with a Lorentzian fit.
The resonance is under-coupled ($\kappa_0>\kappa_\text{ex}$), with fitted $\kappa_0/2\pi=13.8$ MHz. 
Figure \ref{Fig:2}b shows a histogram of $\kappa_0/2\pi$ for 11741 fitted TE$_{00}$ resonances from sixty 100-GHz-FSR microresonators. 
The most probable value is $\kappa_0/2\pi=14$ MHz, corresponding to a statistical intrinsic quality factor of $Q_0=1.4\times10^7$.
The microresonator $Q_0$ and linear optical loss $\alpha$ (dB/m physical length) are linked via  
\begin{equation}
\alpha=27.27\frac{n_g}{\lambda Q_0}
\end{equation} 
At telecommunication band $\lambda=1550$ nm and with a group index $n_g=2.09$ for the given waveguide geometry,  $Q_0=1.4\times10^7$ corresponds to $\alpha=2.6$ dB/m.  
In comparison, $\kappa_0/2\pi=17$ MHz is found for the TM$_{00}$ mode, corresponding to $Q_0=1.1\times10^7$ (see Appendix A).  
The wavelength-dependent loss of each TE$_{00}$ resonance is studied in Appendix B, showing no prominent hydrogen-related absorption around 1520 to 1540 nm. 

Next, we investigate wafer-scale fabrication yield. 
Figure \ref{Fig:2}d right shows our design layout containing sixteen chips on the DUV stepper reticle.
Each chip has a $5\times5$ mm$^2$ size on the wafer and contains many microresonators.  
The DUV stepper exposes the reticle uniformly over the 6-inch wafer in discrete fields. 
The 100-GHz-FSR chips characterized above are labelled as the C11 chips.
The most probable $\kappa_0/2\pi$ values for all C11 chips, as well as their GVD parameters $D_2/2\pi$ (described later), are plotted in each field, as shown in Fig. \ref{Fig:2}d left.
In all measured 20 fields,  $\kappa_0/2\pi\leqslant 16$ MHz is found, demonstrating that our foundry process to manufacture ultralow-loss Si$_3$N$_4$ PIC is uniform and near 100\% yield. 

We also characterize $\kappa_0/2\pi$ of 100-GHz-FSR microresonators of 2.20, 2.00, and 1.80 $\mu$m waveguide width. 
Again, we create a $\kappa_0/2\pi$ histogram for each case and look for the most probable values. 
We plot and compare the most probable $\kappa_0/2\pi$ values for the three width values, and observe a decreasing $\kappa_0/2\pi$ with increasing waveguide width, as shown in Fig. \ref{Fig:2}c. 
This trend indicates that our optical loss is still dominated by the waveguide's Si$_3$N$_4$/SiO$_2$ interface roughness, and can be further reduced by optimizing the dry etching process or using CMP to reduce top surface roughness \cite{Ji:17}. 

\subsection{Microresonator dispersion}
\begin{figure*}[t!]
\centering
\includegraphics[width=13cm]{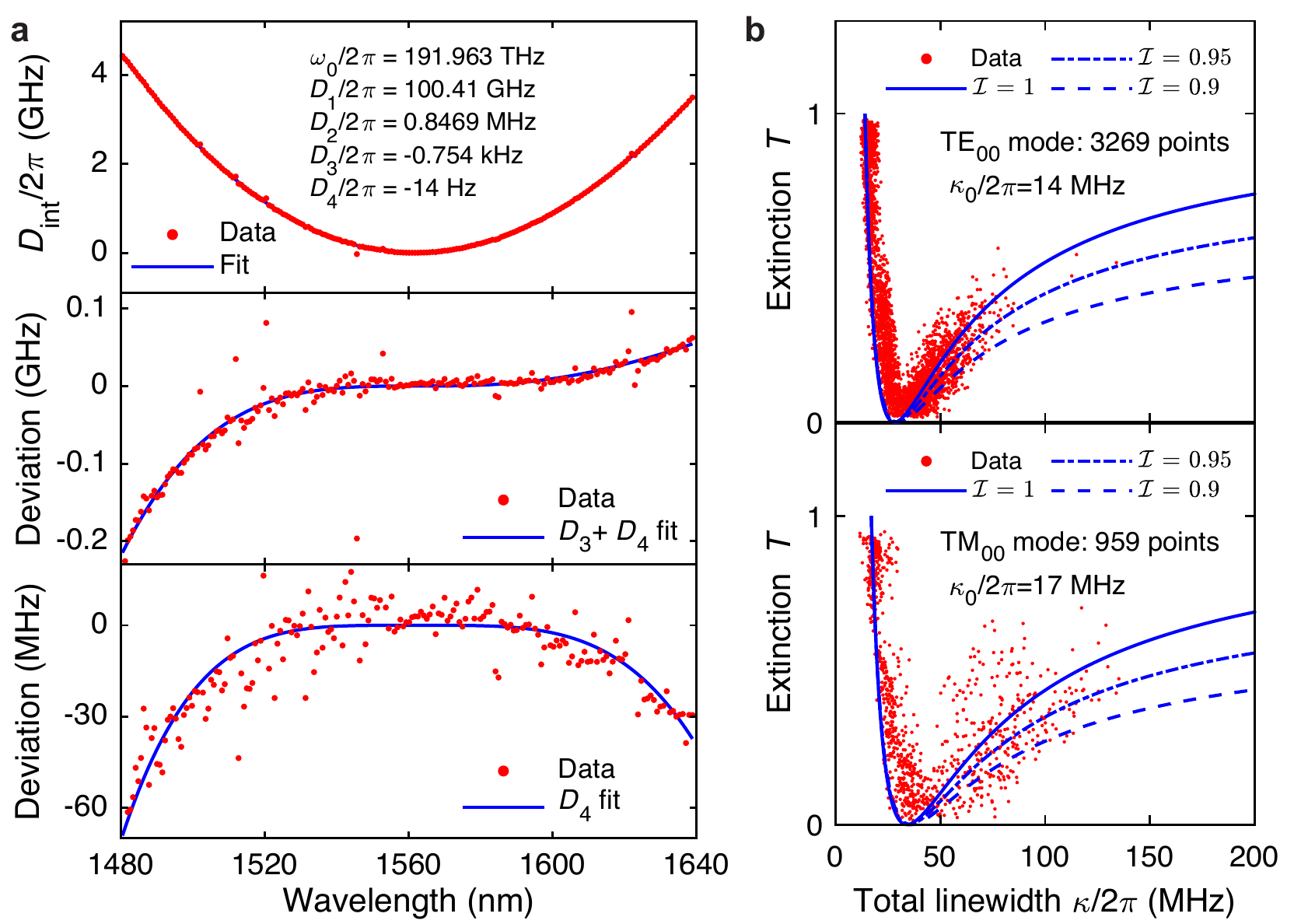}
\caption{
\textbf{Characterization of microresonator dispersion and coupling ideality}.
\textbf{a}.  
Measured integrated dispersion of the microresonator which is fitted with $D_\text{int}(\mu)=D_2\mu^2/2+D_3\mu^3/6+D_4\mu^4/24$ (top), 
the resonance frequency deviations from $D_3\mu^3/6+D_4\mu^4/24$ (middle), 
and the deviations from $D_4\mu^4/24$ (bottom). 
Avoided mode crossings are revealed in the bottom panel, however weak for the later soliton generation experiment. 
\textbf{b}.  
Characterization of coupling ideality of the TE$_{00}$ (top) and TM$_{00}$ (bottom) modes. 
For the TE$_{00}$ / TM$_{00}$ modes, in total thirty-four / seventeen 100-GHz-FSR microresonators are characterized, providing 3269 / 959 data points in each plot. 
A clear trend from under-coupling to critical coupling and then to strong over-coupling is observed. 
The calculated curves of $\mathcal{I}=[0.9, 0.95, 1]$ with $\kappa_0/2\pi=$14 or 17 MHz are plotted for comparison, showing near-unity coupling ideality. 
}
\label{Fig:3}
\end{figure*}

For most applications using Kerr nonlinearity of Si$_3$N$_4$, anomalous GVD is required \cite{Kippenberg:04, Okawachi:14}. 
In addition, for the generation of dissipative Kerr solitons \cite{Herr:14,Brasch:15, Yi:15}, avoided mode crossings (AMX) induced by spatial mode coupling \cite{Herr:14a} should be suppressed, as they prohibit soliton formation \cite{Kordts:16, KimC:21} or distort soliton spectra \cite{Ramelow:14, Huang:16}.
Therefore, next we quantitatively characterize the dispersion profile and investigate AMXs of the high-$Q$, 100-GHz-FSR microresonators. 
The microresonator's integrated dispersion is defined as 
\begin{equation}
\begin{aligned}
D_\text{int}(\mu)&=\omega_{\mu}-\omega_0-D_1\mu\\
&=D_2\mu^2/2+D_3\mu^3/6+D_4\mu^4/24
\end{aligned}
\label{Eq.Dint}
\end{equation}
where $\omega_{\mu}/2\pi$ is the $\mu$-th resonance frequency relative to the reference resonance frequency $\omega_0/2\pi$, 
$D_1/2\pi$ is microresonator FSR, 
$D_2/2\pi$ describes GVD, 
$D_3$ and $D_4$ are higher-order dispersion terms. 
Figure \ref{Fig:3}a top shows a typical $D_\text{int}$ profile, with each parameter extracted from the fit Eq. \ref{Eq.Dint}. 
The positive sign of $D_2$ validates anomalous GVD. 
To reveal AMXs, we remove the $D_2$ term and fit the data with $D_3\mu^3/6+D_4\mu^4/24$, as shown in Fig. \ref{Fig:3}a middle. 
We further remove the $D_2$ and $D_3$ terms, and fit the data with $D_4\mu^4/24$, as shown in Fig. \ref{Fig:3}a bottom. 
The observed AMXs are overall weak and only lead to megahertz-level resonance frequency deviation. 
It also shows that, with these weak AMXs, our spectroscopic method is sufficiently precise to extract the $D_4$ term. 

\subsection{Coupling ideality}
Furthermore, we experimentally characterize the coupling ideality \cite{Spillane:03} of our high-$Q$ Si$_3$N$_4$ microresonators.
In the current case, light in the bus waveguide's fundamental TE mode (TE$_{b, 00}$) is coupled into the microresonator's fundamental TE mode (TE$_{r, 00}$), and then coupled out of the microresonator and back into the bus waveguide. 
In this process, coupling ideality $\mathcal{I}$ is defined as 
\begin{equation}
\mathcal{I}=\frac{\kappa_\text{ex}}{\kappa_\text{ex}+\kappa_p}
\end{equation}
Here $\kappa_\text{ex}$ is the external coupling rate between the TE$_{r, 00}$ and TE$_{b, 00}$ modes,  
$\kappa_p$ is the parasitic loss rate describing coupling strength to other bus waveguide modes as well as radiation modes into free space. 
The parasitic loss $\kappa_p$ appears as another loss channel in addition to $\kappa_0$ and $\kappa_\text{ex}$. 
Thus $\mathcal{I}$ is a parameter describing how much power is recollected in the TE$_{b, 00}$ mode that is exactly the initial driving mode of the bus waveguide \cite{Spillane:03, Pfeiffer:17}.
In the ideal case of single-mode bus waveguide and no radiation into free space, $\mathcal{I}=1$ is obtained. 
In the present case, the bus waveguide is multimode, since it has the same waveguide cross-section (thickness$\times$width) as the microresonator waveguide, to obtain phase matching between the TE$_{b, 00}$ and TE$_{r, 00}$ modes for maximum $\kappa_\text{ex}$. 
Thus the TE$_{r, 00}$ mode can couple to other bus waveguide modes (e.g. higher-order TE modes or any TM modes) or radiation modes, which ultimately compromises coupling ideality ($\mathcal{I}<1$).

With sufficient number of characterized resonances, coupling ideality is evaluated by analyzing the dependence of resonance transmission $T$ and total linewidth $\kappa/2\pi$, as 
\begin{equation}
T=|1-\frac{2}{K^{-1}+\mathcal{I}^{-1}}|^2
\label{Eq.Ideality}
\end{equation}
where $K=\kappa_\text{ex}/\kappa_0$ describes the coupling regime ($K<1$ for under-coupling, $K=1$ for critical coupling, and $K>1$ for over-coupling \cite{Cai:00}). 
Previously, coupling ideality of integrated Si$_3$N$_4$ microresonators has been characterized \cite{Pfeiffer:17}, however based on devices of $Q<4\times10^6$. 
For state-of-the-art Si$_3$N$_4$ microresonators of $Q>10^7$, coupling ideality has not been experimentally studied, and whether high coupling ideality is still maintained needs to be answered. 

Here we perform extra measurement on over-coupled devices with smaller gap values down to 300 nm. 
The microresonators are identical, except that the gap varies from sample to sample to provide a varying $\kappa_\text{ex}$. 
Figure \ref{Fig:3}b shows the measured coupling ideality for the TE$_{00}$ and TM$_{00}$ modes from dozens of 100-GHz-FSR microresonators. 
The TE$_{00}$ (top) and TM$_{00}$ (bottom) plots contain 3269 and 959 data points, respectively. 
The data is sufficient to uncover the global trend even in the presence of $\kappa_0$ variation due to fabrication and AMXs.
Meanwhile, for comparison, we also plot the $\kappa$-$T$ curves with $\mathcal{I}=[0.9,0.95, 1]$ calculated using Eq. \ref{Eq.Ideality}. 
As can be seen, our microresonators coupled with curved bus waveguides can already provide near-unity $\mathcal{I}$ and strong over-coupling (e.g. $K>7$ for the TM$_{00}$), which are critical for microring-based phase modulation \cite{Bogaerts:12, Liang:21},  wideband tunable delay line \cite{Xiang:18}, and the extraction of quantum light states generated in high-$Q$ microresonators \cite{Vernon:16, Dutt:15, Vaidya:20, Perez:23}. 

\section{Soliton microcomb generation}
\begin{figure*}[t!]
\centering
\includegraphics[width=13cm]{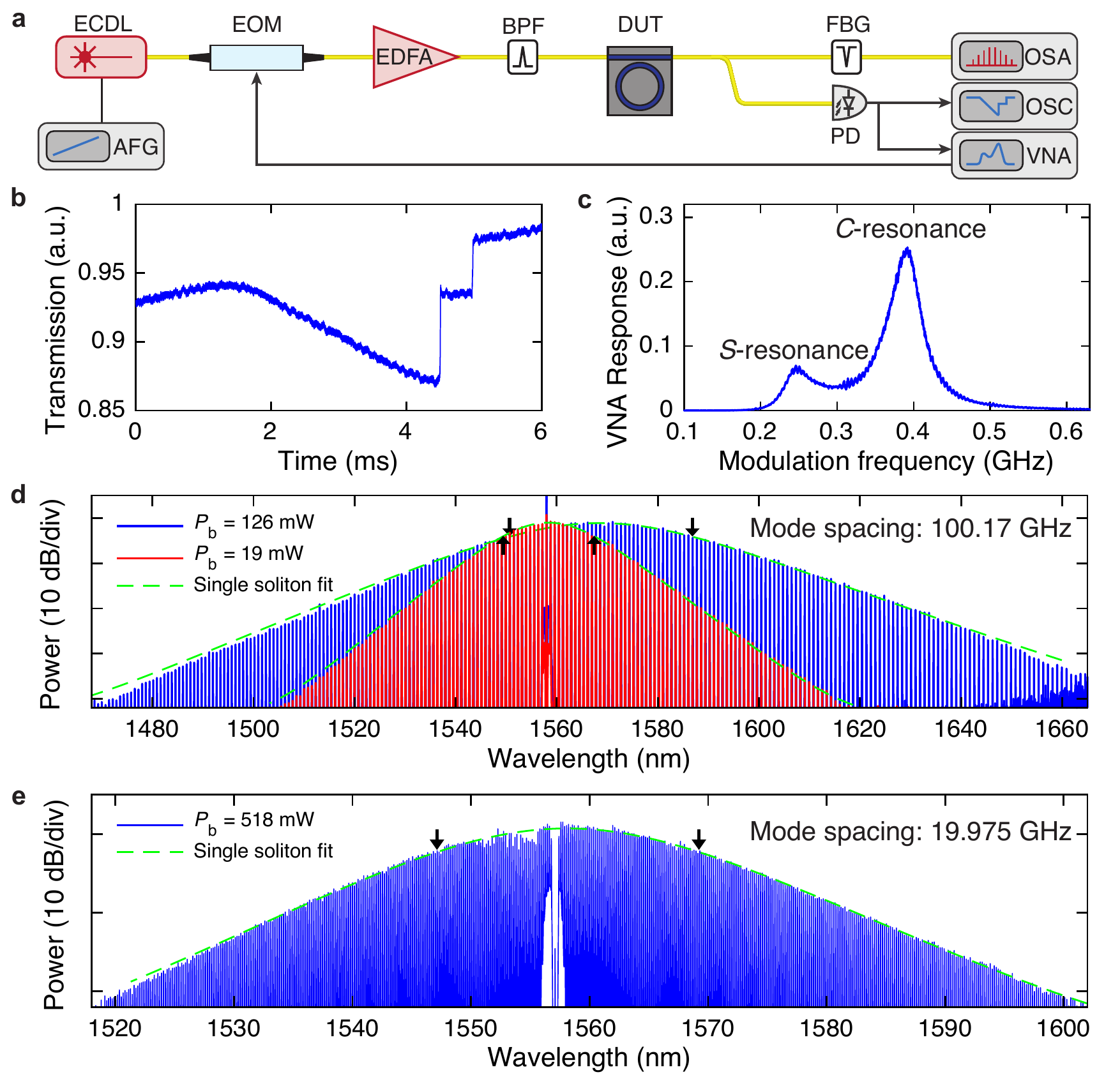}
\caption{
\textbf{Single-soliton generation in silicon nitride}.
\textbf{a}. 
Experimental setup. 
AFG, arbitrary function generator; 
ECDL,external-cavity diode laser; 
EDFA, erbium-doped fiber amplifier; 
BPF, bandpass filter; 
EOM, electro-optic modulator;
DUT, device under test;
FBG, fiber Bragg grating; 
PD, photodiode;
OSA, optical spectrum analyser; 
OSC, oscilloscope; 
VNA, vector network analyzer. 
\textbf{b}.  
When the laser frequency is scanned from the blue-detuned to the red-detuned side of a resonance, a soliton step of sub-millisecond length appears,enabling direct access to soliton states via simple piezo tuning of laser frequency. 
\textbf{c}. 
Cavity response measurement using the EOM and VNA. 
The appearance of $\mathcal{S}$-resonance verifies soliton generation. 
\textbf{d}. 
Single-soliton spectra of 100.17 GHz mode spacing, with 19 mW (red) and 126 mW (blue) CW pump power on the chip, and their spectral fit (green). 
With 19 / 126 mW power, the arrows mark the 3-dB bandwidth of 17.99 / 35.68 nm. 
A prominent Raman self-frequency shift of 10.4 nm is observed with 126 mW power.
\textbf{e}. 
Single-soliton spectrum of 19.975 GHz mode spacing with 518 mW (blue) CW pump power on the chip, and its spectral fit (green). 
The arrows mark the 3-dB bandwidth of 21.96 nm, containing 137 comb lines. 
In both \textbf{d} and \textbf{e}, the EDFA’s amplified spontaneous emission (ASE) noise is filtered out by the BPF, and the pump laser in the soliton spectra is filtered out by the FBG. 
}
\label{Fig:4}
\end{figure*}
A key application area of our Si$_3$N$_4$ PIC is Kerr nonlinear photonics, where ultralow optical loss is central as it determines the threshold power $P_\text{th}$ for Kerr parametric oscillation \cite{Kippenberg:04}. 
For example, for soliton microcomb generation \cite{Herr:14, Brasch:15, Yi:15}, $P_\text{th}$ scales with the microresonator $Q$ factor as $P_\text{th}\propto Q^{-2}$.
Therefore, with this quadratic dependence, a high $Q$ enables significant reduction of $P_\text{th}$ down to milliwatt level. 

The experimental setup to generate single solitons in our Si$_3$N$_4$ chips is shown in Fig. \ref{Fig:4}a. 
We note that, due to the high-$Q$ and low thermal effects of our Si$_3$N$_4$ microresonators, single solitons can be generated via simple laser-piezo frequency tuning\cite{Herr:14, Guo:16} or on-chip heaters \cite{Joshi:16, Xue:16}.
This is in contrast to many soliton generation experiments that require sophisticated techniques to manage thermal effects \cite{Li:17} such as power kicking \cite{Brasch:16, Yi:16b}, single-sideband suppressed-carrier frequency shifters \cite{Stone:18}, dual-laser pump \cite{Zhou:19, ZhangS:19}, pump modulation \cite{Wildi:19,Nishimoto:22}, pulse pumping \cite{Obrzud:17}, or laser self-cooling \cite{Lei:22}.  

When the continuous-wave (CW) pump laser's frequency scans across a resonance from the blue-detuned side to the red-detuned side, a step feature (i.e. the ``soliton step'') is observed in the microresonator transmission spectrum \cite{Herr:14}, signalling soliton formation.  
Figure \ref{Fig:4}b shows a typical soliton step with sub-millisecond duration, sufficiently long for accessing the single-soliton state via simple laser-piezo frequency tuning \cite{Herr:14, Guo:16}. 
To further confirm the soliton nature and measure the soliton detuning value, a system response measurement using an electro-optic modulator (EOM) and a vector network analyzer (VNA) is performed \cite{Guo:16}. 
As shown in Fig. \ref{Fig:4}c, the system response features double resonances corresponding to the cavity resonance of the CW pump (``$\mathcal{C}$-resonance''), and the soliton-induced resonance (``$\mathcal{S}$-resonance''). 
Physically, the ``cold'' cavity resonance is probed by the intracavity low-power CW as a background to the soliton pulse pattern, thus the observed $\mathcal{C}$-resonance frequency indicates the effective laser-cavity detuning (where thermal induced resonance shift is eliminated) \cite{Guo:16}. 
The $\mathcal{S}$-resonance is induced by the soliton pattern that has a high peak-power leading to nonlinear frequency shift of the cavity resonance.

As shown in Fig.~\ref{Fig:4}d, in a 100-GHz-FSR microresonator,  a single soliton of 100.17 GHz mode spacing is generated with 19 mW power in the bus waveguide on-chip. 
The spectrum sech$^2$ fit shows 3-dB bandwidth of 17.99 nm, corresponding to a Fourier-limited pulse duration of 141.8 fs. 
Increasing the pump power to 126 mW and the pump laser detuning, the soliton's 3-dB spectrum bandwidth is increased to 35.68 nm (pulse duration of 71.47 fs). 
We also observe a strong Raman-induced self-frequency shift \cite{Karpov:16, Yi:16} of 10.4 nm. 

Moreover, we generate a single soliton in a 20-GHz-FSR microresonator from the same wafer.  
Figure~\ref{Fig:4}e shows the single soliton spectrum of 19.975 GHz mode spacing with 518 mW power in the bus waveguide.  
The 3-dB bandwidth of 21.96 nm, corresponding to a pulse duration of 115.4 fs, covers 117 comb lines. 
This coherent soliton microcomb with a microwave K-band repetition rate is advantageous for applications such as high-spectra-efficiency telecommunications \cite{Mazur:21, Fujii:22}, photonic microwave generation \cite{Liang:15, Liu:20, Yang:21}, and astronomical spectrometer calibration \cite{Suh:19, Obrzud:19}. 

Previously, among all CMOS-compatible high-index materials, single-soliton microcombs of repetition rates below microwave K-band ($<20$ GHz) have only been realized in Si$_3$N$_4$, using either the 4-inch photonic Damascene process \cite{Liu:20}, or the 3-inch EBL-written subtractive process \cite{Ye:22}.  
Our work represents the first foundry-based, 6-inch subtractive process with DUV stepper lithography to reach this goal. 

\section{Conclusion and outlook}
In conclusion, we have reported a 6-inch foundry fabrication process of tight-confinement, dispersion-engineered Si$_3$N$_4$ PIC of optical loss down to 2.6 dB/m and near 100\% yield. 
We have demonstrated its application in soliton microcomb generation with low power threshold and dense channel spacings.
While currently our process is based on 6-inch wafers due to our dry etcher, essentially our process can be scaled up to an 8-inch process, which can offer even better uniformity and higher throughput. 
Merging our ultralow-loss Si$_3$N$_4$ process with established heterogeneous integration \cite{Park:20, opdeBeeck:20} can introduce a variety of active functions to the passive Si$_3$N$_4$ PIC, such as narrow-linewidth lasers in the UV and visible band \cite{WangZ:17, Corato-Zanarella:22}, broadband EOMs \cite{WangC:18, Snigirev:21}, fast photodetectors \cite{Yu:20, Lin:22}, and programmable MEMS-controlled network \cite{Dong:22, Tian:20}. 
Together, foundry development of heterogeneous, ultralow-loss Si$_3$N$_4$ integrated photonics could revolutionize next-generation applications for frequency metrology \cite{Spencer:18, Newman:19}, photonic neural networks \cite{Bogaerts:20, Shastri:21}, and quantum information processing \cite{WangJW:20, Mehta:20}.

\section*{Appendix A: Characterization of the microresonator TM$_{00}$}
\begin{figure}[t!]
\centering
\includegraphics[width=13cm]{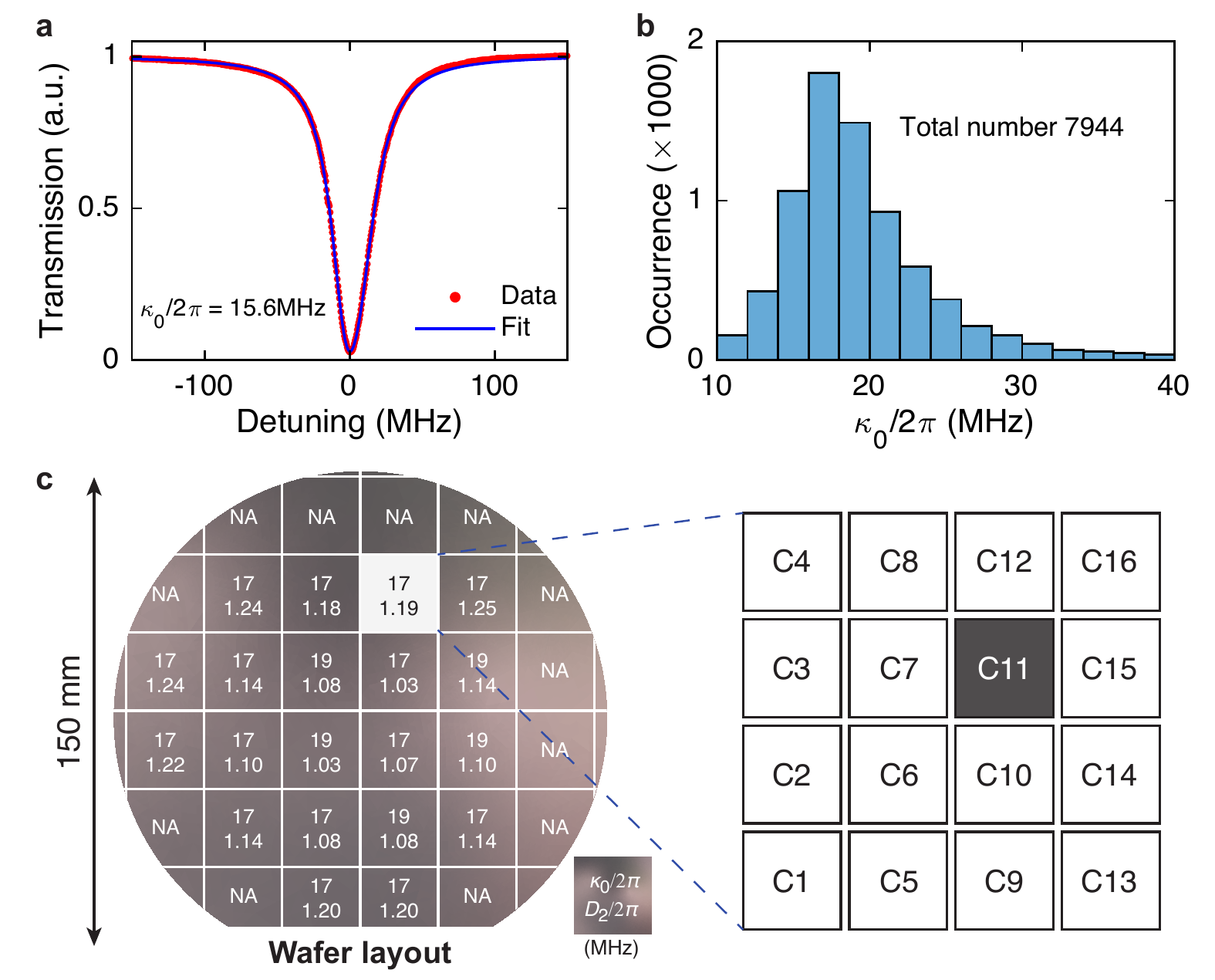}
\caption{
\textbf{Characterization of the microresonator TM$_{00}$ mode}. 
\textbf{a}.  
A typical TM$_{00}$ resonance profile with a Lorentzian fit, showing $\kappa_0/2\pi=15.6$ MHz. 
\textbf{b}.  
Histogram of 7944 TM$_{00}$ resonances from forty 100-GHz-FSR microresonators of 2.40 $\mu$m waveguide width, showing the most probable value of $\kappa_0/2\pi=17$ MHz and $Q_0=1.1\times10^7$.
\textbf{c}.  
Uniformity and yield analysis over the 6-inch wafer scale. 
Right: The DUV stepper reticle layout contains sixteen chips, and is uniformly exposed in discrete fields over the 6-inch wafer.
Left: The most probable values $\kappa_0/2\pi$ of the C11 chips, as well as the measured GVD parameters $D_2/2\pi$, are marked in each fields over the wafer.
NA: not applicable, due to visible defects or missing C11 chips near wafer edge.
}
\label{SIfig:1}
\end{figure}

Figure \ref{SIfig:1}a shows a TM$_{00}$ resonance with a Lorentzian fit.
The resonance is critically coupled ($\kappa_0\approx\kappa_\text{ex}$), with fitted $\kappa_0/2\pi=15.6$ MHz. 
Figure \ref{SIfig:1}b shows a histogram of $\kappa_0/2\pi$ for 7944 fitted TM$_{00}$ resonances from forty 100-GHz-FSR microresonators. 
The most probable value is $\kappa_0/2\pi=17$ MHz, corresponding to $Q_0=1.1\times10^7$.
Figure \ref{SIfig:1}c shows the most probable $\kappa_0/2\pi$ values for all C11 chips plotted in each field of the 6-inch wafer, as well as their GVD parameters $D_2/2\pi$. 
In all measured 20 fields,  $\kappa_0/2\pi\leqslant 19$ MHz is found. 

\section*{Appendix B: Loss versus wavelength}
\begin{figure}[t!]
\centering
\includegraphics[width=8cm]{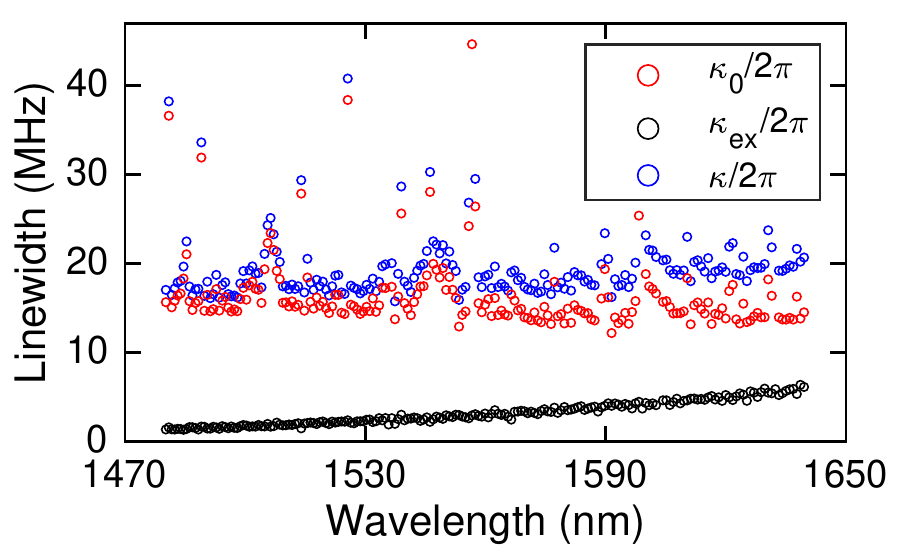}
\caption{
\textbf{Broadband measurement of resonance linewidth}.  
Measured and fitted $\kappa_0/2\pi$, $\kappa_\text{ex}/2\pi$, and $\kappa/2\pi=(\kappa_0+\kappa_\text{ex})/2\pi$ of each resonance from 1480 to 1640 nm. 
The alignment of $\kappa_\text{ex}/2\pi$ values on a line indicates correct resonance fit with reasonable precision.
Local $\kappa_0/2\pi$ increase at multiple wavelengths are likely caused by AMXs. 
No prominent hydrogen-related absorption around 1520 to 1540 nm is observed. 
}
\label{SIfig:2}
\end{figure}

We use frequency-comb-assisted diode laser spectroscopy\cite{DelHaye:09} to measure resonance frequency $\omega/2\pi$ and linewidth $\kappa/2\pi$, ranging from 1480 to 1640 nm wavelength. 
Figure \ref{SIfig:2} shows the measured and fitted intrinsic loss $\kappa_0/2\pi$, the external coupling strength $\kappa_\text{ex}/2\pi$, and the total (loaded) linewidth $\kappa/2\pi=(\kappa_0+\kappa_\text{ex})/2\pi$ of each resonance of a typical 100-GHz-FSR microresonator.  
Since the bus waveguide and the microresonator are coupled via evanescent field, $\kappa_\text{ex}/2\pi$ is wavelength-dependent with a given geometry. 
Therefore the alignment of $\kappa_\text{ex}/2\pi$ values on a line indicates correct resonance fit with reasonable precision. 
Local $\kappa_0/2\pi$ increase is observed at multiple wavelengths, however such narrow-band features are likely caused by AMXs. 
In addition, no prominent hydrogen-related absorption around 1520 nm to 1540 nm is observed, indicating low photo-thermal absorption.

\section*{Appendix C: CMP dishing effect}

\begin{figure}[b!]
\centering
\includegraphics[width=13cm]{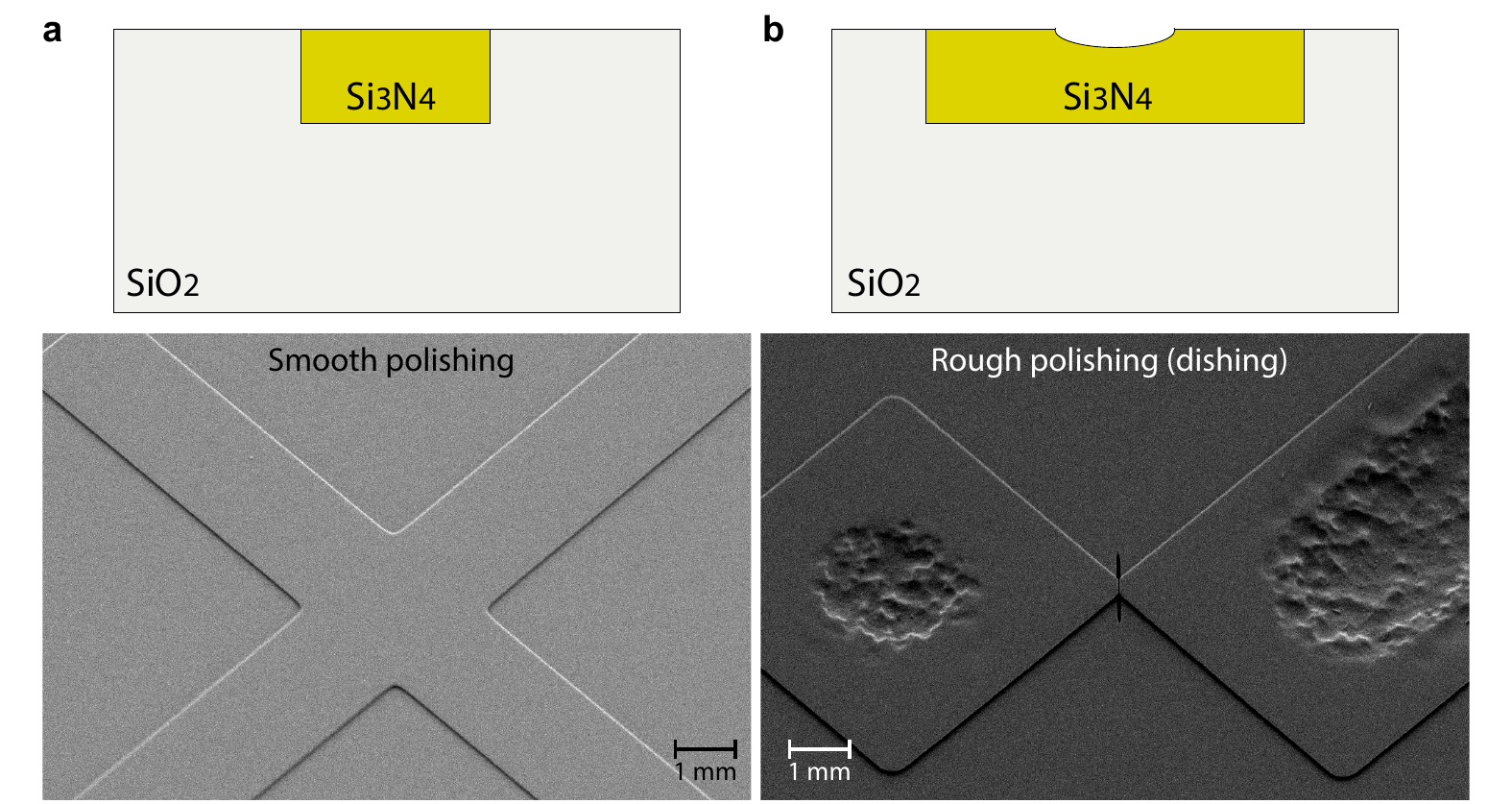}
\caption{
\textbf{Illustration of the CMP dishing effect observed during fabrication.} 
\textbf{a}.
SEM images showing smooth top surface after CMP for waveguides with smaller critical dimension (e.g. below 3 $\mu$m). 
\textbf{b}.
SEM images showing rough top surface due to the CMP dishing effect for waveguides with larger critical dimension (above 3 $\mu$m). 
}
\label{SIfig:3}
\end{figure}

Our reported process is based on the subtractive process that is a top-down process where Si$_3$N$_4$ waveguides are formed by dry etching. 
There is another process widely used for the fabrication of ultralow-loss Si$_3$N$_4$ PIC, i.e. the photonic Damascene process \cite{Pfeiffer:18b, Liu:21}. 
This process is an ``additive'' process. 
As illustrated in Ref. \cite{Liu:21}, the patterns are transferred from the photoresist mask to the SiO$_2$ substrate to create waveguide preforms. 
Then an LPCVD Si$_3$N$_4$ film is deposited on the patterned substrate, filling the preform trenches and forming the waveguides. 
Chemical-mechanical planarization (CMP) is used to remove excess Si$_3$N$_4$ and create a flat and smooth wafer top surface.  
The rest steps are the same as the subtractive process. 

It should be noted that the dishing effect, illustrated in Fig. \ref{SIfig:3}, is commonly presented if the CMP polishing rates for the waveguide material and cladding are different (which is true for Si$_3$N$_4$ and SiO$_2$).  
In the Damascene case, the CMP slurry containing SiO$_2$ nano-particles causes the polishing rate of Si$_3$N$_4$ higher than that of thermal wet SiO$_2$, which induces the dishing effect in large-area patterns (e.g. critical dimension larger than 3 $\mu$m). 
The dishing effect leads to significant structure distortion and top surface roughness.


\begin{backmatter}
\bmsection{Funding}
J. Liu acknowledges support from the National Natural Science Foundation of China (Grant No.12261131503), 
Hetao Shenzhen-Hong Kong Science and Technology Innovation Cooperation Zone Project (No. HZQB-KCZYB-2020050), 
and from the Guangdong Provincial Key Laboratory (2019B121203002).
Y.-H L. acknowledges support from the China Postdoctoral Science Foundation (Grant No. 2022M721482). 

\bmsection{Acknowledgments}
J. Liu is indebted to Dapeng Yu who provided critical support to this project. 
We thank Chao Xiang for the fruitful discussion.  

\bmsection{Author contributions}
J. Liu and Z. Y.  conceived the fabrication process. 
Z. Y., H. J. and Z. H. developed the fabrication process and fabricated the samples. 
C. S. took the SEM images and provided support in the fabrication.   
J. Long, B. S., Y.-H. L. and W. S. built the experiment and characterized the samples, while H. G., J. H. and Z. Y. provided support and suggestion. 
W. S. , J. Long and J. H. performed the soliton generation experiment. 
L. G. took the photos of the sample. 
Y.-H. L, B. S.  and J. Liu analysed the data and prepared the manuscript with input from others. 
J. Liu supervised the project and managed the collaboration. 

\bmsection{Disclosures}
Z. Y., H. J., Z. H. and J. Liu are co-founders of Qaleido Photonics, a start-up that is developing heterogeneous silicon nitride integrated photonics technologies.  
Others declare no conflicts of interest. 

\bmsection{Data Availability Statement}
are available on Zenodo (https://doi.org/10.5281/zenodo.7639970). 
All other data used in this study are available from the corresponding authors upon reasonable request.

\bmsection{Additional information}
This manuscript has been published on Photonics Research (doi:10.1364/PR.486379). 
We encourage the readers to cite our formal publication on Photonics Research.
\end{backmatter}




\end{document}